\documentclass[12pt]{article}
\usepackage{amsmath}
\usepackage{amssymb}
\usepackage{graphicx}
\usepackage{ifthen}
\usepackage{verbatim}
\usepackage{psfrag}
\usepackage{color}

\numberwithin{equation}{section}

\newcommand{\asi}{\,\!}

\begin{document}

\bigskip
\bigskip
\begin{center}
{\Large \textbf{Coadjoint Orbits\\
 and\\
 Wilson Loops\\
 in\\
 Five Dimensional Topological Gauge Theories
}}
\end{center}

\vspace{10mm}
\begin{center}
Yui Noma
\footnote{E-mail: \texttt{yuhii@mpim-bonn.mpg.de}}
\\
\bigskip
{\small 
\textit{
Max-Planck-Institut f\"{u}r Mathematik\\
Bonn, Germany\\}}
\end{center}

\vspace{10mm}
\begin{abstract}


We discussed
 a one-point function
 of a BPS Wilson loop
 in a supersymmetric five dimensional
 gauge theories
 defined on $M_4\times S^1$
 by using 
 path integral expression
 of Wilson loops.
We found that
 the Wilson loop
 gives interaction terms between
 charged particles and certain gauge fields
 on the instanton moduli space,
 and
 makes the non-charged particle charged
 under the gauge fields.

\end{abstract}
\newpage

\section{Introduction}

We discussed
 one-point functions
 of BPS Wilson loops
 in a five dimensional topological
 gauge theory
 defined on $M_4\times S^1$:
\begin{eqnarray}
  \mathrm{Tr}_{V_\lambda}
  \left(
   \mathrm{P}
   \exp\left(
	-\sqrt{-1}
	\int_0^T dt
	(A_t+\sqrt{-1}\varphi)^I X_I 
       \right)
  \right)
  .
\end{eqnarray}
This Wilson loops are important observable
 as studied in \cite{nakatsu} and \cite{nakatsu2}
 for the non-commutative $U(1)$ gauge theory.
It is difficult to discuss
 correlation functions
 of Wilson loops
 because it is cumbersome
 to
 treat
 the path-ordering operator,
 which is needed to define Wilson loops.
In order to remove this complexity,
 we used a path integral expression
 of Wilson loops.
The expression
 is an extension 
 of relations between
 Kirillov`s formula of characters of a Lie group
 and
 quantum mechanics  for  particles
 whose phase spaces are coadjoint orbits.
According to this path integral expression,
 the path-ordering operator in a Wilson loop
 is considered as the time-ordering operator
 of the quantum mechanics.

It is known  \cite{5d and sqm}
 that
 by using the supersymmetry
 the partition function of the 
 five dimensional topological gauge theory
 is expressed as
 a partition function
 of a quantum mechanics of
 a non-charged particle moving
 on a instanton moduli space.
Furthermore,
 the quantum mechanics possesses
 a supersymmetry.
Due to the supersymmetry,
 the path integral variables of 
 the partition function of the quantum mechanics
 are localized to constant modes
 along the circle $S^1$.
As a result, the partition function
 of the quantum mechanics 
 becomes the index of the Dirac operator
 of the instanton moduli space
 \cite{5d and sqm}.
Because the Wilson loops are BPS with respect
 to the supersymmetry of the topological gauge theory,
 this argument are applicable
 to correlation functions 
 of the Wilson loops.
After the variables are localized to
 the constant modes,
 the Wilson loops are considered
 as Chern characters of some vector bundles,
 and their one-point functions
 become the indices of the twisted Dirac operators
 of the vector bundles.
Although 
 it is known that indices of twisted Dirac operators
 are given as partition functions
 of quantum mechanics of charged particles
 \cite{charged particle and index},
 it is not known a direct relation
 between the Wilson loops
 and charged particles.
By using the path integral expression
 of Wilson loops,
 we found that
 the Wilson loop
 gives interaction terms between
 a charged particle and a certain gauge field
 on the instanton moduli space,
 and
 makes the non-charged particle charged
 under the gauge fields.
\begin{eqnarray}
 \langle
  \mathrm{Tr}_{V_\lambda}
  \left(
   \mathrm{P}
   \exp\left(
	-\sqrt{-1}\int_0^T dt
	(A_t+\sqrt{-1}\varphi)^I X_I 
       \right)
  \right)
  \rangle
  \nonumber\\
  =
  \int_{\mathcal{M}\times\Omega_{\lambda+\rho}}
  \mathcal{D}m^i
  \mathcal{D}\chi^i
  \mathcal{D}\phi^a
  \mathcal{D}c^a
  \, e^{-T \tau \int_{M_4}c_2(AdP)}
  e^{-S_{charged}}  
.
\nonumber\\
\end{eqnarray}

This paper
 is organized as follows.
In section \ref{sec;wilson and coad},
 we introduce coadjoint orbits
 and relations between
 coadjoint orbits and Wilson loops.
In section \ref{sec;coad and 5d},
 we apply the method argued in section \ref{sec;wilson and coad}
 to Wilson loops in five dimensional topological gauge theories.
We denoted geometries of spaces of connections
 in appendix \ref{sec;appendix}
 in order to complement the discussion in the main part of this paper.

\section{Wilson loops and coadjoint orbits}
\label{sec;wilson and coad}

This section is a preparation
 for the discussion in the section \ref{sec;coad and 5d}
 and do not possess any new results.
Firstly, we introduce
 coadjoint orbits in 
 \ref{subsec;coad and Kirillov formula}. 
In particular, the most important equation
 is the Kirillov formula
 (\ref{eq;Kirillov formula}).
Secondly,
 we denote a relation
 between characters
 and quantum mechanics on
 coadjoint orbits
 in \ref{subsec;coad and SQM}.
Thirdly, we give a short derivation
 of the path integral expression
 of 
 Wilson loops
 in \ref{subsec;wilson and QM}.

\subsection{Coadjoint orbit and Kirillov's formula}
\label{subsec;coad and Kirillov formula}
In this subsection, 
 we explain coadjoint actions
 and coadjoint orbits.
For details of the things in this subsection,
 see Kirillov's lecture note \cite{Kirillov}
 and the references in the lecture note.
We take elements of Lie algebras
 as Hermitian matrices
 in order to clarify a relation
 between 
 coadjoint orbits and
 quantum mechanics,
 which is denoted in the later subsections.

Let $G$ be a real simple compact Lie group,
 and $\mathfrak{g}$ be the Lie algebra of $G$.
We take $\{X_I\}_{I=1,\cdots,\dim{\mathfrak{g}}}$
 as a basis of the Lie algebra $\mathfrak{g}$.
The algebra is given by the relations
 $[X_I,X_J]=\sqrt{-1}f_{IJ}\asi^K X_K$,
 where $f_{IJ}\asi^K$
 is the structure constant.
The Cartan subalgebra is denoted by $\mathfrak{h}$.
We write the corresponding maximal torus as $H$:
 $Lie(H)=\mathfrak{h}$.

The Lie group $G$ acts on $\mathfrak{g}$.
The action is denoted by $(\mathfrak{g}, Ad)$
 and called the adjoint representation.
The infinitesimal action of $G$ gives
 the action of Lie albegra $\mathfrak{g}$
 on $\mathfrak{g}$.
The action is denoted by $(\mathfrak{g},ad)$
 and also called the adjoint representation:
\begin{eqnarray}
 ad &:& \mathfrak{g} \rightarrow End(\mathfrak{g}),\\
 ad(X)(Y)&:=&
 \left[X,Y\right],
\hspace{5mm}\mbox{for $X,Y \in \mathfrak{g}$}.
\end{eqnarray}

Let $\mathfrak{g}^*$ be the dual vector space of $\mathfrak{g}$.
The pairing is denoted by $\langle\, ,\,\rangle$:
\begin{eqnarray}
 \langle\, ,\,\rangle &:&
  \mathfrak{g}^* \times \mathfrak{g}
  \rightarrow \mathbb{R}.
\end{eqnarray}
We define the coadjoint representation 
 $(\mathfrak{g}^*,K)$ of $G$ as the dual representation 
 of the adjoint representation $(\mathfrak{g}, Ad)$.
In the same way,
 the coadjoint representation $(\mathfrak{g}^*,k)$
 of $\mathfrak{g}$
 as the dual representation of $(\mathfrak{g}, ad)$:
\begin{eqnarray}
 \langle K(g)F,Y\rangle
  &=&
  \langle F,Ad(g^{-1})Y\rangle,
  \\
 \langle k(X)F,Y\rangle
  &=&
  - \langle F,ad(X)Y\rangle,
  \hspace{5mm}\mbox{for 
  $g\in G$,
  $X,Y \in \mathfrak{g}$,
  $F\in \mathfrak{g}^*$}.  
\end{eqnarray}

A $G$-orbit passing through $F_0\in\mathfrak{g}^*$
 is called a coadjoint orbit of $G$ and
 is denoted by $\Omega_{F_0}$:
\begin{eqnarray}
 \Omega_{F_0}
  &:=&
  \left\{
   F \in \mathfrak{g}
   |
   F= K(g)F_0,
   \hspace{2mm}
   ^\exists g\in G 
  \right\}.
\end{eqnarray}
From now on, 
 when the coadjoint orbit under consideration
 is definite,
 we omit the subscript and denote the coadjoint orbit 
 by $\Omega$.
Let $Stab(F)\subset G$ be the stabilizer group 
 of the point $F\in\mathfrak{g}$:
\begin{eqnarray}
 Stab(F)&:=&
  \left\{
   g\in G
   |
   K(g)F=F
  \right\}.
\end{eqnarray}
The coadjoint orbit $\Omega_{F_0}$ 
 is diffeomorphic
 to the homogeneous manifold
 $G/Stab(F_0)$:
\[
 \Omega_{F_0}
 \simeq
 G/Stab(F_0).
\]

We denote the coordinate of $\mathfrak{g}^*$ 
 by $\{q_I\}_{I=1,\cdots,dim \mathfrak{g}}$,
 and
 the coordinate of $\Omega$
 by $\{\phi^a\}_{a=1,\cdots,dim\Omega}$.
An element $X\in \mathfrak{g}$ acts on
 a coadjoint orbit $\Omega$,
 and generates a vector field
 $X_{\Omega}\in \Gamma(\Omega,T\Omega)$.
By means of the coordinate,
 the vector field $X_{\Omega}$ is 
 given as follows:
\begin{eqnarray}
 X_{\Omega}
  |_F
  &=&
  -\sqrt{-1}
  \langle k(X)F, X_I \rangle
  \frac{\partial \phi^a}
  {\partial q_I}
  \frac{\partial}
  {\partial \phi^a},
  \hspace{5mm}\mbox{at $F\in \Omega$.}
  \label{eq;X_Omega}
\end{eqnarray}

For each coadjoint orbit $\Omega$,
 there is a symplectic form $\omega$
 defined by the following equation:
\begin{eqnarray}
 \omega(X_{\Omega},Y_{\Omega})(F)
  &:=&
  \sqrt{-1}
  \langle F, \left[X,Y\right]
  \rangle,
  \label{eq;symplectic form}
\end{eqnarray}
where $X_{\Omega}$, $Y_{\Omega}$ are 
 the vector fields on $\Omega$
 generated by the elements $X,Y\in \mathfrak{g}$.
The symplectic form $\omega$ is 
 called the Kirillov-Kostant two-form.
The symplectic form $\omega$ is $G$-invariant.
Hence, it vanishes by the action of the Lie derivative;
 $\mathcal{L}(X_{\Omega})(\omega)=0$
 for any element $X\in\mathfrak{g}$.

Let $V_X$ be the Hamilton vector field
 generated by the function 
 $X\in\mathfrak{g}$ on $\Omega$:
\begin{eqnarray}
 V_X
  :=
  \omega^{ab}\partial_b X\partial_a.
\end{eqnarray}
Since the inverse matrix of the symplectic form
 is given by the following equation,
\begin{eqnarray}
 \omega^{ab}(F)
  =
  f_{IJ}^K 
 \langle F, X_K \rangle
  \frac{\partial \phi^a}{\partial q_I}
  \frac{\partial \phi^b}{\partial q_J}
  ,
\end{eqnarray}
the vectors $V_{X_I}$ and $X_{I\Omega}$ are same:
\begin{eqnarray}
 V_{X_I}|_F
  &=&
  f_{JK}^L 
  \langle F, X_K \rangle
  \frac{\partial \phi^a}{\partial q_J}
  \frac{\partial \phi^b}{\partial q_K}
  \frac{\partial q_I}{\partial \phi^b}
  \partial_a
  \nonumber\\
 &=&
  \sqrt{-1}\langle F, 
  [X_I,X_J]\rangle 
  \frac{\partial \phi^a}{\partial q_J}
  \partial_a
  \nonumber\\
 &=&
  X_{I\Omega}|_F.
\end{eqnarray}
In general, $V_{X}=X_{\Omega}$
 for any $X\in\mathfrak{g}$.

Every coadjoint orbit is a K\"{a}hler
 manifold.
Then,
 there is a $G$-invariant metric $g_{\Omega}$
 on each coadjoint orbit $\Omega$.

Let $\mathfrak{g}_{\mathbb{C}}:=
\mathfrak{g}\otimes_\mathbb{R}\mathbb{C}$
 be the complexification of $\mathfrak{g}$.
We denote the Cartan subalgebra of 
$\mathfrak{g}_{\mathbb{C}}$ by $\mathfrak{h}_{\mathbb{C}}$.
Let $\lambda\in \mathfrak{h}_{\mathbb{C}}^*$
 be a dominant weight.
The coadjoint orbit $\Omega_{\lambda}$
 is integral.
That is, the symplectic form 
 satisfies the following conditions:
\begin{eqnarray}
 \int_{C_2} \omega
  \in 2\pi \mathbb{Z},
  \hspace{5mm}
  \mbox{for any two cycles } C_2 \in H_2(\Omega_{\lambda},\mathbb{Z}). 
\end{eqnarray}
If the dominant weight $\lambda$
 is in the open Weyl chamber,
 the coadjoint orbit $\Omega_{\lambda}$
 is diffeomorphic to the 
 full flag manifold $G/H$.

The following expression of characters
 is the one of the most important relation
 in this paper.
Let $(\pi_\lambda, V_\lambda)$ be
 the highest weight representation of $\mathfrak{g}$
 with the highest weight $\lambda$.
For an element $e^{-\sqrt{-1}X}\in G$ which is 
 sufficiently close to the identity element,
 the character of the element
 is given as follows by using Kirillov's formula:
\begin{eqnarray}
 \mathrm{Tr}_{V_\lambda}
  \left(
   e^{-\sqrt{-1}X}
  \right)
  =
  \frac{1}{p(X)^{1/2}}
  \int_{F\in\Omega_{\lambda+\rho}}
  e^{-\sqrt{-1}\langle F,X\rangle}
  e^{\frac{\omega}{2\pi}},
  \label{eq;Kirillov formula}
\end{eqnarray}
where $\rho$ is the half of the sum of the positive roots
 of $\mathfrak{g}_\mathbb{C}$
 and $p(X)$ is defined by the following equation:
\begin{eqnarray}
 p(X)
  :=
  \det
  \left(
   \frac{\sinh\left(ad(\sqrt{-1}X)/2\right)}
   {\left(ad(\sqrt{-1}X)/2\right)}
  \right).
\end{eqnarray}

\subsection{Quantum mechanics on coadjoint orbit
 and Kirillov's formula}
\label{subsec;coad and SQM}


A partition function
 of a certain quantum mechanics
 gives a character
 of a Lie group:
\begin{eqnarray}
 \mathrm{Tr}_{V_\lambda}
  \left(
   e^{-\sqrt{-1}T X_I}
  \right)
  &=&
  Z_{\Omega}
,
  \label{eq;char eq part}
  \\
 Z_{\Omega}
  &:=&
  \int \mathcal{D}\phi \mathcal{D}c
  \,
  e^{\sqrt{-1}S_\Omega}
  ,
\end{eqnarray}
where $X_I\in\mathfrak{h}$ and
$T$ is the circumference
 of $S^1$ which is the time direction of the
 quantum mechanics.
$S_{\Omega}$ 
 is an action whose Hamiltonian is $X_I$.
The real function $X_I$ on the coadjoint orbit
 becomes a Hermitian matrix after the quantization.
This is the reason why we took elements
 of Lie algebras as Hermitian matrices.
We will describe the details of the quantum mechanics
 later.
The relation eq.(\ref{eq;char eq part})
 is denoted in \cite{char and SQM}
 and be proved
 by localizing
 the both hand side.
We give a short proof of the relation.

The integrand of eq.(\ref{eq;Kirillov formula})
 is a equivariantly closed
 differential form.
\begin{eqnarray}
 \left(
  d-2\pi \sqrt{-1}\iota(T V_{X_I})
 \right)
 \left(
  \frac{\omega}{2\pi}
  -\sqrt{-1}T X_I
 \right)
 =
 0.
\end{eqnarray}
Then,
 we can apply the localization formula
 for integrals of equivariantly closed forms
 to
 the integration
 in Kirillov's formula.
The integration is determined by the information
 around the zero locus of
 the vector field $V_{X_I}$.
We already know that
 the vector field $X_{I\Omega}$
 and
 $V_{X_I}$ are same.
Hence, the zero locus of the
 vector field $V_{X_I}$
 is the fixed points
 of the action of $X_I\in\mathfrak{g}$.
According to the following discussion,
 we will know
 that
 the fixed points are isolated
 on the coadjoint orbit $\Omega_{\lambda+\rho}$.
Let $F\in\Omega_{\lambda+\rho}$ be a fixed point
 of $X_I$-action.
We assume that
 there exist an element $Y\in\mathfrak{g}$
 such that the point
 $K(e^{\sqrt{-1}Y})F\in\Omega_{\lambda+\rho}$
 is also a fixed point of $X_I$-action.
With this assumption,
 because of 
 $K(e^{-\sqrt{-1}Y}e^{\sqrt{-1}X_I} e^{\sqrt{-1}Y})F =F$,
 the element
 $e^{-\sqrt{-1}Y}e^{X_I} e^{\sqrt{-1}Y}$
 is in the stabilizer of $F$:
 $e^{-\sqrt{-1}Y}e^{X_I} e^{\sqrt{-1}Y}\in Stab(F)=\mathfrak{h}$.
Then, $e^{\sqrt{-1}Y}$
 is in the normalizer subgroup $N_G(H)$
 of $H$ in $G$.
Since $N_G(H)/H$ is the Weyl group
 and the Weyl group is a discrete group,
 $Y\in\mathfrak{h}$
 if $Y$ is close to the origin.
Therefore, 
 $K(e^{\sqrt{-1}Y})F= F$
 and
 the fixed points are isolated.
In this case,
 Kirillov's formula
 becomes as follows:
\begin{eqnarray}
 \mathrm{Tr}_{V_\lambda}
  \left(
   e^{-\sqrt{-1}TX_I}
  \right)
  &=&
  \frac{1}{p(TX_I)^{1/2}}
  \left(
   -\sqrt{-1}
  \right)^{(dim \mathfrak{g}-rank\mathfrak{g})/2}
  \sum_{
  \substack{p\in\Omega_{\lambda+\rho}\\
 X_{I\Omega}|_p=0}
  }
  \frac{e^{-\sqrt{-1}TX_I}|_p}
  {\det_{ab}\left(
	     T\partial_b X_{I\Omega}^a|_p\right)^{1/2}}
  .
  \label{eq;localized Kirillov formula}
\end{eqnarray}

The partition function
 of the quantum mechanics
 is localized
 by using the method
 denoted in \cite{SQM part exact calc}.
$S_{\Omega}$
 is given as follows:
\begin{eqnarray}
S_\Omega
 &:=&
 \int_0^T dt
 \left(
 \theta_a \dot{\phi}^a
 -X_I
 +\frac{1}{2}
 \omega_{ab}
 c^a c^b
\right).
\end{eqnarray}
The field contents of the quantum mechanics
 are
 real scalar fields $\phi^a$,
 which correspond to a coordinate
 of $\Omega_{\lambda+\rho}$,
 and
 real fermion fields $c^a$,
 which correspond to 
 the differential forms $d\phi^a$.
Furthermore,
 we have used a symplectic
 potential $\theta_a$
 for the symplectic form $\omega$:
\begin{eqnarray}
 d\theta = \omega.
\end{eqnarray}
The action $S_{\Omega}$
 is symmetric
 with respect to
 the following
 supersymmetry
 transformation:
\begin{eqnarray}
 Q_\Omega \phi^a
  &=&
  c^a,
  \\
 Q_\Omega c^a
  &=&
  \dot{\phi}^a
   -V_{X_I}^a
  .
\end{eqnarray}
Since the action is $Q_\Omega$-closed,
 the partition function
 $Z_{\Omega}$ is independent
 of deformations
 of the action with terms
 which are not only
 $Q_\Omega$-exact but also $Q_\Omega$-closed.
By deforming the action,
 the partition function
 is localized
 as follows:
\begin{eqnarray}
 Z_{\Omega}
  =
  \mathcal{N}
  \sum_{
  \substack{p\in\Omega_{\lambda+\rho}\\
 V_{X_I}|_p=0}
  }
  \det_{ab}
  \left(
   \frac{T\partial_a V_{X_I}^b/2}
   {\sinh\left(T\partial_a V_{X_I}^b/2\right)}
  \right)^{1/2}
  \hspace{-5mm}(p)
  \frac{e^{-\sqrt{-1}TX_I(p)}}
  {\det_{ab}\left(
	     T\partial_a V_{X_I}|_p
	    \right)^{1/2}}
  ,
  \label{eq;localized part of omega}
\end{eqnarray}
where $\mathcal{N}$ is a normalization factor.
By comparing
 eq.(\ref{eq;localized part of omega})
 with eq.(\ref{eq;localized Kirillov formula}),
 we recognize that
 the equation (\ref{eq;char eq part}) holds
 if $\mathcal{N}=(-\sqrt{-1})^{dim\mathfrak{g}-rank\mathfrak{g}}$
 and the first determinant factor in
 eq.(\ref{eq;localized part of omega})
 is equal to
 $p(-TX_i)^{-1/2}$.
The determinant in  $p(-TX_i)^{-1/2}$
 is taken over
 $\mathfrak{g}$.
Because the adjoint action
 of an element in  $\mathfrak{h}$
 on $\mathfrak{h}$
 is trivial,
 the space over which
 the determinant in  $p(-TX_i)^{-1/2}$
 is taken 
 is reduced to
 $\mathfrak{g}/\mathfrak{h}$.
Furthermore by using eq.(\ref{eq;X_Omega}),
 $p(-TX_i)^{-1/2}$
 becomes
 the first determinant factor in
 eq.(\ref{eq;localized part of omega}).

\subsection{Path integral expression of Wilson loops}
\label{subsec;wilson and QM}

The aim of this subsection
 is that
 we write Wilson loops of gauge theories
 as path integrals of certain quantum mechanics
 by extending the correspondence between
 characters and partition functions
 (\ref{eq;char eq part}).
This is an old idea
 started from
 \ref{Witten}.
The key ideas in this subsection
 are
 as follows:
 the path-ordering operator in a Wilson loop 
 is considered as 
 the time-ordering operator of a quantum mechanics
 and
 the non-commutativity of the Lie algebra
 is regarded as a quantum effect.

We consider a gauge theory on a manifold $M$.
The gauge group is denoted by $G$. 
The coordinate of $M$ is denoted by
 $x^\mu$.
Let $A_{\mu}$ be the gauge field.
For a given loop $\gamma:[0,T]\rightarrow M$
 and a given highest representation
 $(\pi_\lambda,V_\lambda)$ of $G$,
 a Wilson loop is defined by the following equation:
\begin{eqnarray}
 W(\gamma,\lambda)
  :=
  \mathrm{Tr}_{V_\lambda}
  \left(
   \mathrm{P}
   \exp
   \left(
    -\sqrt{-1}
    \int_0^T dt A_\mu^I \frac{dx^\mu}{dt}
    X_I
   \right)
  \right)
  .
\end{eqnarray}
If the gauge field is constant along the loop,
 the Wilson loop $ W(\gamma,\lambda)$
 can be written
 as a partition function.
Hence, we decompose 
 the gauge field
 into a constant part
 and the others:
\begin{eqnarray}
 A_\mu^I \dot{x}^\mu
(\gamma(t))
  =
  A_{0\mu}^I\dot{x}^\mu
  +
  A_{t\mu}^I\dot{x}^\mu(\gamma(t))
  .
\end{eqnarray}
With this decomposition,
 the Wilson loop is expanded as follows:
\begin{eqnarray}
  W(\gamma,\lambda)
  &=&
  \mathrm{Tr}_{V_\lambda}
  \left(
   U(T,0)
  \right)
  \nonumber\\
 &&
  +
  \int_0^T dt_1
  \mathrm{Tr}_{V_\lambda}
  \left(
   U(T,t_1)
   \mathcal{O}(t_1)
   U(t_1,0)
  \right)
  \nonumber\\
 &&
  +
  \int_0^T dt_1
  \int_0^{t_1} dt_2
  \mathrm{Tr}_{V_\lambda}
  \left(
   U(T,t_1)
   \mathcal{O}(t_1)
   U(t_1,t_2)
   \mathcal{O}(t_2)
   U(t_2,0)
  \right)
  \nonumber\\
  &&+\cdots
  ,
  \label{eq;expand of Wilson}
\end{eqnarray}
where
\begin{eqnarray}
 U(t_1,t_2)
  &:=&
  \exp
  \left(
   {-\sqrt{-1}(t_1-t_2)A_{0\mu}^I \dot{x}^\mu X_I}
  \right)
  ,
  \\
 \mathcal{O}(t)
  &:=&
  \left(
   -\sqrt{-1}A_{t\mu}^I \dot{x}^\mu X_I
  \right)(\gamma(t))
  .
\end{eqnarray}
The first term in eq.(\ref{eq;expand of Wilson})
 is written as the partition function
 of quantum mechanics whose
 Hamiltonian is
 $A_{0\mu}^I \dot{x}^\mu X_I(\phi)$.
The integrand of the second term
 is a correlation function
 of a operator $\mathcal{O}(t_1)$
 inserted in the time $t_1$.
The other terms
 are also considered
 as integrals of 
 correlation functions
 of operators $\mathcal{O}(t_i)$.
The $r+1$-th term
 is written as follows:
\begin{eqnarray}
 &&\int_0^T dt_1 
  \int_0^{t_1}dt_2
  \cdots
  \int_0^{t_{r-1}}dt_r
 \mathrm{Tr}_{V_\lambda}
  \left(
   U(T,t_1)\mathcal{O}(t_1)
   U(t_1,t_2)\mathcal{O}(t_2)
   \cdots
   U(t_{r-1},t_r)\mathcal{O}(t_r)
   U(t_r,0)
  \right)
  \nonumber\\
 &&
  =
  \int_0^T dt_1 
  \int_0^{t_1}dt_2
  \cdots
  \int_0^{t_{r-1}}dt_r
  \int_{\Omega_{\lambda+\rho}}
  \mathcal{D}\phi\mathcal{D}c
  \left(
   \prod_{i=1}^r
   \mathcal{O}(t_i)
  \right)
  e^{\sqrt{-1}S_\Omega}
  .
  \label{eq;r+1 term 1}
\end{eqnarray}
Since
 operators are treated as
 classical quantities  in a path integral,
 for any pairs of time $t_i$ and $t_j$
 the operators $\mathcal{O}(t_i)$
 and $\mathcal{O}(t_j)$
 commute
 in the path integral.
Then eq.(\ref{eq;r+1 term 1})
 is simplified as follows:
\begin{eqnarray}
  \int_0^T dt_1 
  \int_0^{t_1}dt_2
  \cdots
  \int_0^{t_{r-1}}dt_r
  \int_{\Omega_{\lambda+\rho}}
  \mathcal{D}\phi\mathcal{D}c
  \left(
   \prod_{i=1}^r
   \mathcal{O}(t_i)
  \right)
  e^{\sqrt{-1}S_\Omega}
 \nonumber\\
 =
  \frac{1}{r!}
  \int_0^T dt_1
  \int_0^T dt_2
  \cdots
  \int_0^T dt_r
  \int_{\Omega_{\lambda+\rho}}
  \mathcal{D}\phi\mathcal{D}c
  \left(
   \prod_{i=1}^r
   \mathcal{O}(t_i)
  \right)
  e^{\sqrt{-1}S_\Omega}
  .
\end{eqnarray}
By applying the above argument,
 we arrive at the following path integral expression
 of the Wilson loop:
\begin{eqnarray}
 W(\gamma,\lambda)
  &=&
  \int_{\Omega_{\lambda+\rho}}
  \mathcal{D}\phi\mathcal{D}c
  \sum_{r=0}^\infty
  \left(
   \frac{1}{r!}
   \prod_{i=1}^r
   \int_0^T dt_i
   \mathcal{O}(t_i)
  \right)
  e^{\sqrt{-1}S_\Omega}  
  \\
 &=&
  \int_{\Omega_{\lambda+\rho}}
  \mathcal{D}\phi\mathcal{D}c
  \,
  e^{\sqrt{-1}\tilde{S}_\Omega}  
  ,
  \label{eq;wilson as sqm}
\end{eqnarray}
where $\tilde{S}_{\Omega}$
 is an action with a time-dependent Hamiltonian
 $A_\mu^I\dot{x}^\mu(\gamma(t))  X_I(\phi)$
:
\begin{eqnarray}
 \tilde{S}_{\Omega}
  :=
  \int_0^T dt
  \left(
   \theta_a \dot{\phi}^a
   -A_\mu^I\dot{x}^\mu(\gamma(t))
   X_I(\phi(t))
   +\frac{1}{2}
   \omega_{ab}
   c^a c^b
  \right).
\end{eqnarray}
The equation (\ref{eq;wilson as sqm})
 is nothing but
 the equalities
 of partition functions 
 of quantum mechanics
 with a time-dependent Hamiltonian.

\section{Coadjoint orbits
 and Wilson loops in 5D supersymmetric gauge theories}
\label{sec;coad and 5d}

In this section,
 we apply the relation
 between Wilson loops and quantum mechanics
 to Wilson loops
 of certain
 supersymmetric
 five dimensional gauge theories
 defined on a five dimensional manifold $M_4\times S^1$.
The Wilson loops which we will consider
 are the following
 ones wrapping around the circle in the fifth direction:
\begin{eqnarray}
  \mathrm{Tr}_{V_\lambda}
  \left(
   \mathrm{P}
   \exp\left(
	-\sqrt{-1}\int_0^T dt
	(A_t+\sqrt{-1}\varphi)^I X_I 
       \right)
  \right)
  ,
\end{eqnarray}
where $A_t$ is the fifth component
 of the gauge field
 and $\varphi$
 is a scalar field.
The five dimensional gauge theories which we will consider
 is
 five dimensional topological non-abelian gauge theories
 defined on a Cartesian product
 of a general four dimensional manifold
 $M_4$ 
 and a circle $S^1$.
The minimally coupled $\mathcal{N}=1$
 supersymmetric five dimensional
 gauge theory
 defined on a flat manifold $\mathbb{R}^4\times S^1$
 is one of the five dimensional topological gauge theories.

It is known that
 the four dimensional component
 of the action 
 of the topological gauge theories
 can be integrated
 \cite{5d and sqm}.
The resulting action
 is an action
 of a particle moving on
 instanton moduli space $\mathcal{M}$.
The coordinate $t$
 if the fifth direction is
 treated
 as a time variable of the quantum mechanics.
The particle is not charged under any gauge groups.
We will show that
 the Wilson loops
 give interaction terms
 of charged particles 
 and gauge fields.
Therefore,
 we showed
 a direct relation
 between
 Wilson loops
 and charged particles.

At first we describe shortly the known relation
 between the five dimensional gauge theory
 and a particle moving on instanton moduli
 space $\mathcal{M}$
 in subsection \ref{subsec;5d and sqm}.
In subsection \ref{subsec;wilson and charged particle},
 we will show that
 the interaction terms of a charged particle
 and gauge field
 are coming from the Wilson loop.

\subsection{5D supersymmetric gauge theory and
quantum mechanics on instanton moduli space}
\label{subsec;5d and sqm}


The minimally coupled $\mathcal{N}=1$ supersymmetric
 five dimensional gauge
 theory 
 is treated as a topological gauge theory
 by the following method.
We introduce the standard flat metric
 on the five dimensional Euclidean manifold
$\mathbb{R}^4\times S^1$. 
The coordinate of $S^1$
 is denoted by $t$.
The circumference of the circle $S^1$ is $T$.
By regarding the $\mathcal{N}=1$ supersymmetric gauge theory
 as a $\mathcal{N}=2$ supersymmetric four dimensional
 gauge theory depending on $t$,
 we twist the theory
 like the Donaldson-Witten theory. 
By this procedure
 the theory can be seen 
 as a five dimensional topological gauge theory.

We denote a coordinate of $M_4$ by $x^\mu$.
The field contents of the topological gauge theory
 is as follows:
 a gauge field $A_\mu$ and $A_t$,
 an adjoint real scalar field $\varphi$,
 a real scalar fermion field $\eta$,
 a real one-form fermion field $\psi_\mu dx^\mu$
 and a real self-dual fermion field
 $\frac{1}{2}\xi_{\mu\nu} dx^\mu\wedge dx^\nu$.
Here we have defined the self-duality
 by using the Hodge star operator * defined 
 on the four dimensional manifold $M_4$.
There is a scalar supercharge $Q_{YM}$.
Because the supersymmetry transformations
 caused by $Q_{YM}$
 are not closed on shell,
 we add an auxiliary
 self-dual field
 $H_{\mu\nu}$
 to close the supersymmetry transformations
 on shell.
With the auxiliary field,
 $(A,\psi,A_t+\sqrt{-1}\varphi)$,
 $(A_t-\sqrt{-1}\varphi,\eta)$
 and $(\xi_{\mu\nu},H_{\mu,\nu})$
 form multiplets
 under the supercharge $Q_{YM}$.
The most important multiplet
 is $(A,\psi,A_t+\sqrt{-1}\varphi)$
 whose supersymmetry
 transformations are:
\begin{eqnarray}
& Q_{YM} A
  =
  \psi
  ,
  \hspace{5mm}
 Q_{YM} \psi
  =
  d_A(A_t+\sqrt{-1}\varphi)
  -\partial_t A
  ,
&  \\
& Q_{YM} (A_t+\sqrt{-1}\varphi)
  =
  0
  .
  \label{eq;supertrans of SYM}
\end{eqnarray}
We have used a covariant derivative operator
 $d_A:=d+A^I ad(\sqrt{-1}X_I)$ acting on $\mathfrak{g}$-valued
 differential forms.
The action consists of
 the standard action $S_{YM}$
 and the following topological term $S_{top}$:
\begin{eqnarray}
S_{top}
 =
 \frac{\theta  \sqrt{-1}}{T}
  \int_{S^1}dt
  \left(\frac{-1}{8\pi^2}\right)
  \int_{M_4}
  \mathrm{Tr}_{\mathfrak{g}}
  \left(
   F\wedge F
  \right)
  ,
\end{eqnarray}
where $\theta$ is a coupling constant.
The standard action $S_{YM}$ can be written as
 a $Q_{YM}$-exact term up to
 a topological term:
\begin{eqnarray}
 S_{YM}+S_{top}
  =
  \int_{M_4\times S^1}
  \left\{
   Q,W
  \right\}
  +
  \tau
  \int_{S^1}dt
  \left(\frac{-1}{8\pi^2}\right)
  \int_{M_4}
  \mathrm{Tr}_{\mathfrak{g}}
  \left(
   F\wedge F
  \right)
  ,
\end{eqnarray}
 where
 $\tau:=   \frac{\theta  \sqrt{-1}}{T}+\frac{4\pi^2}{g_{YM}^2}$.
Due to this property of the action,
 correlation functions of BPS observables
 are independent of
 deformations
 of the action with terms
 which are not only
 $Q_{YM}$-exact but also $Q_{YM}$-closed.
From now on, to make expressions simple
 we set $g_{YM}=1$.
We deform the action except the topological term
 as follows:
\begin{eqnarray}
 \int_{M_4\times S^1}
  d^5x
  \mathrm{Tr}_{\mathfrak{g}}
  \left\{
   Q_{YM},
   F_{\mu\nu}\xi^{\mu\nu}
   -\frac{1}{2}
   \psi^\mu
   (D_\mu(A_t-\sqrt{-1}\varphi)-\partial_t A_\mu)
  \right\}
  .
  \label{eq;5d henkeigo}
\end{eqnarray}
With this deformed action,
 the equations of motions 
 of the multiplets
 $(A_t-\sqrt{-1}\varphi,\eta)$
 and
 $(\xi_{\mu\nu},H_{\mu\nu})$
 give restrictions
 on the multiplet
$(A,\psi,A_t+\sqrt{-1}\varphi)$.
The equations of motion
 of the multiplet
 $(\xi_{\mu\nu},H_{\mu\nu})$
 are as follows: 
\begin{eqnarray}
 F^{(+)}=0,
  \hspace{5mm}
  (d_A\psi)^{(+)}=0,
  \label{eq;eom 1}
\end{eqnarray}
where the symbol $^{(+)}$ means the self-dual component.
According to this conditions,
 the domain of the path integral
 is restricted to
 the moduli space of anti-instantons
 and tangent vectors on them. 
By integrating out
 $(A_t-\sqrt{-1}\varphi,\eta)$, 
 we obtain the following equations:
\begin{eqnarray}
 &d_A^*\psi=0,
  \label{eq;eom 2}
  \\
 &  
  d_A^* d_A (A_t+\sqrt{1}\varphi)
  -d_A^* \partial_t A
  -2\sqrt{-1}\psi\wedge\psi
  =0
  .
  \label{eq;phi to kyokuritu}
\end{eqnarray}
We have used the adjoint operator
 $d_A^*$ of $d_A$:
 $d_A^*=-*d_A*$ for $\mathfrak{g}$-valued one-forms.

After the integration of the multiplets,
 the deformed action becomes as follows:
\begin{eqnarray}
 \int
  \mathrm{Tr}_{\mathfrak{g}}
  \left(
   \frac{1}{2}
   \partial_t A^\mu
   (D_\mu(A_t+\sqrt{-1}\varphi)-\partial_t A_\mu)
   -\frac{1}{2}
   \psi^\mu \partial_t \psi_\mu
  \right)
  \label{eq,action remaining 1}
  .
\end{eqnarray}
The four dimensional part
 of the remaining action (\ref{eq,action remaining 1})
 can be integrated out
 by using eq.(\ref{eq;del_i A}) and (\ref{eq;Gij}).
The result is
 the following action $S_{SQM}$
 for a particle moving
 on the instanton moduli space $\mathcal{M}$:
\begin{eqnarray}
 S_{SQM}
  &:=&
  \int_0^T dt
  \hspace{1mm}
  \frac{1}{2}G_{ij}\dot{m}^i\dot{m}^j
  +\frac{1}{2}
  \chi^i
  \left(
   G_{ij}\partial_t
   +\dot{m}^k
   G_{il}\Gamma^l_{kj}
  \right)
  \chi^j
  .
\end{eqnarray}
In order to derive the above action,
 we have used the following equation
 for the Levi-Civita connection
 $\Gamma^k_{jl}$:
\begin{eqnarray}
G_{ik}\Gamma^k_{jl}(m)
 =
 \int_{M_4}
 \mathrm{Tr}_{\mathfrak{g}}
 \left(
  b_i\wedge * \nabla_j b_l
 \right)
 ,
\end{eqnarray}
where
 $\nabla_i := \partial_i + \epsilon_i^I ad(\sqrt{-1} X_I)$
 is the $\mathcal{A}^*/\mathcal{G}_0$
 direction of the covariant derivative
 of the universal bundle $\mathcal{E}$.
See Appendix \ref{sec;appendix} for details.

\subsection{One-point function of Wilson loop
 and quantum mechanics of charged particle}
\label{subsec;wilson and charged particle}

We discuss a one-point function
 of a Wilson loop
 and show that
 the Wilson loop
 gives the interaction terms
 of a charged particle and
 a gauge field.

Let $\gamma_{x_0}:[0,T]\rightarrow M_4\times S^1$
 be a path on $M_4\times S^1$
 such that
$\gamma_{x_0}(t)=(x_0,t)\in M_4\times S^1$ for $t\in[0,T]$.
For a given highest representation
 $(\pi_\lambda,V_\lambda)$,
 we define the following
 Wilson loop:
\begin{eqnarray}
 W(\gamma_{x_0},\lambda)
  =
  \mathrm{Tr}_{V_\lambda}
  \left(
   \mathrm{P}
   \exp\left(
	-\sqrt{-1}\int_0^T dt
	(A_t+\sqrt{-1}\varphi)^I X_I 
       \right)
  \right)
  .
\end{eqnarray}
According to the equation (\ref{eq;wilson as sqm}),
 we write the Wilson loop
 $W(\gamma_{x_0},\lambda)$
 as a partition function
 of a quantum mechanics on a coadjoint orbit:
\begin{eqnarray}
 W(\gamma_{x_0},\lambda)
 =
 \int_{\Omega_{\lambda+\rho}}
 \mathcal{D}\phi^a
 \mathcal{D}c^a
 \exp
 \left(
  \sqrt{-1}S_{Wilson}
 \right)
 ,
\end{eqnarray}
where
\begin{eqnarray}
 S_{Wilson}
  &:=&
  \int_0^T dt\,
  \theta_a \dot{\phi}^a
  -(A_t+\sqrt{-1}\varphi)^I
  X_I
  +\frac{1}{2}\omega_{ab}
  c^a c^b
  .
\end{eqnarray}

Because of the supersymmetry transformations (\ref{eq;supertrans of SYM}),
 the Wilson loop
 $W(\gamma_{x_0},\lambda)$
 is BPS:
\begin{eqnarray}
 Q_{YM}
  W(\gamma_{x_0},\lambda)=0.
\end{eqnarray}
Furthermore, 
 the Wilson loop
 consists of $A_t+\sqrt{-1}\varphi$,
 which was not integrated out in subsection \ref{subsec;5d and sqm}.
Therefore,
 we can apply the argument in
 subsection \ref{subsec;5d and sqm}.
By combining the previous argument
 and the path integral expression
 of the Wilson loop,
 the one-point function
 of the Wilson loop
 becomes as follows:
\begin{eqnarray}
 &&
  \langle
  W(\gamma_{x_0},\lambda)
  \rangle
  \\
 &=&
  \int
  \mathcal{D}\phi_{YM}
  e^{-S_{top}-S_{YM}}
  \mathrm{Tr}_{V_\lambda}
  \left(
   \mathrm{P}
   \exp\left(
	-\sqrt{-1}\int_0^T dt
	(A_t+\sqrt{-1}\varphi)^I X_I 
       \right)
  \right)
  \\
  &=&
 \int_{\mathcal{M}}
 \mathcal{D}m^i
 \mathcal{D}\chi^i
 \int_{\Omega_{\lambda+\rho}}
 \mathcal{D}\phi^a
 \mathcal{D}c^a
 \exp
 \left(
  -\tau
  \int_{S^1}dt
  \left(\frac{-1}{8\pi^2}\right)
  \int_{M_4}
  \mathrm{Tr}_{\mathfrak{g}}
  \left(
   F\wedge F
  \right)
  -S_{SQM}
  +\sqrt{-1}S_{Wilson}
 \right)
 .
 \label{eq;one point func of wilson 1}
 \nonumber\\
\end{eqnarray}
Hence, the effect of the Wilson loop
 $W(\gamma_{x_0},\lambda)$
 is concentrated
 on the Hamiltonian
 $(A_t+\sqrt{-1}\varphi)^I(x_0,m,\chi) X_I(\phi)$.

We solve $(A_t+\sqrt{-1}\varphi)^I$
 as a function
 of $m^i$ and $\chi^i$.
It is known that $(A_t+\sqrt{-1}\varphi)^I$
 is related to the curvature of the universal bundle
 \cite{5d and sqm}.
At first, we solve $\psi_\mu$
 as a function
 of $m^i$ and $\chi^i$.
The equations for $\psi_\mu$
 (\ref{eq;eom 1}) and (\ref{eq;eom 2})
 are solved by using
 $b_i$ defined in eq.(\ref{eq;def of bi}):
\begin{eqnarray}
 \psi_\mu= b_{i\mu}\chi^i
  .
\end{eqnarray}
By comparing eq.(\ref{eq;phi to kyokuritu})
 with
 eq.(\ref{eq;curv of univ}),
 $A_t+\sqrt{-1}\varphi$ 
is solved as follows:
\begin{eqnarray}
 A_t+\sqrt{-1}\varphi
  =
  \dot{m}^i\epsilon_i
  -\frac{1}{2}
  F_{ij}\chi^i\chi^j
  .
  \label{eq;phi to kyokuritu 2}
\end{eqnarray}

By substituting
 eq.(\ref{eq;phi to kyokuritu 2}) into
 eq.(\ref{eq;one point func of wilson 1}),
 the one-point function
 of the Wilson loop
 becomes
 as follows:
\begin{eqnarray}
  \langle
  W(\gamma_{x_0},\lambda)
  \rangle
  =
 \int_{\mathcal{M}\times\Omega_{\lambda+\rho}}
 \mathcal{D}m^i
 \mathcal{D}\chi^i
 \mathcal{D}\phi^a
 \mathcal{D}c^a
 \,
 e^{-T \tau \int_{M_4}c_2(AdP)}
 e^{-S_{charged}}
,
\end{eqnarray}
 where
 $c_2(AdP)$ is the second Chern class
 of the vector bundle $AdP$ defined in \ref{subsec;geom of a/g}
 and $S_{charged}$ is defined as follows:
\begin{eqnarray}
 S_{charged}
  &:=&
  \int_0^T dt 
  \hspace{1mm}
  \frac{1}{2}G_{ij}\dot{m}^i\dot{m}^j
  -\sqrt{-1}\theta_a\phi^a
  +\dot{m}^i
  \epsilon_i^I(x_0,m) \sqrt{-1}X_I
  \\
 &&
  +\frac{1}{2}
  \chi^i
  \left(
   G_{ij}\partial_t
   +\dot{m}^k
   G_{il}\Gamma^l_{kj}
   -F_{ij}(x_0,m)\chi^i\chi^j \sqrt{-1}X_I
  \right)
  \chi^j
  -\frac{\sqrt{-1}}{2}
  \omega_{ab}c^a c^b
  .
\end{eqnarray}
$S_{charged}$
 is nothing but 
 an action for a charged particle
 denoted in \cite{charged particle and index}
 except the imaginary unit $\sqrt{-1}$.
The reason of the appearance of 
 the imaginary unit is
 the following difference
 of the treatment of $S^1$:
 in one hand
 $S^1$ is considered 
 as a part of the Euclidean space
 in the gauge theory,
 in the other hand
 $S^1$
 is regarded as the time of
 the quantum mechanics on the coadjoint orbit.

There is a general theory
 that a partition funciton
 of a charged particle
 gives 
 a index of
 a twisted Dirac operator
 of
 a vector bundle
\cite{charged particle and index}.
To use the theory,
 we need to determine the vector bundle
 under consideration.
The charged particle which we are considering
 is a particle moving
 on the instanton moduli space $\mathcal{M}$
 and is charged
 with respect to $G$
 as the representation $\pi_{\lambda}$.
Furthermore,
 the local connection one-form $\epsilon_i$
 and the curvature $F_{ij}$
 are
 evaluated at
 a fixed point $x_0\in M_4$.
Hence, the appropriate
 vector bundle
 is 
 $\iota_{x_0}^*( \mathcal{E}_{\lambda})$,
 which is defined
 in subsection \ref{subsec;vector bundle on a/g},
 restricted to the instanton moduli space
 $\mathcal{M}$:
 $\iota_{x_0}^* (\mathcal{E}_{\lambda})|_{\mathcal{M}}$.
Therefore,
 we arrive at the following fact
 that
 the one-point function
 of the Wilson loop
 is 
 the index of 
 the vector bundle
$\iota_{x_0}^* (\mathcal{E}_{\lambda})|_{\mathcal{M}}$:
\begin{eqnarray}
 \langle
  \mathrm{Tr}_{V_\lambda}
  \left(
   \mathrm{P}
   \exp\left(
	-\sqrt{-1}\int_0^T dt
	(A_t+\sqrt{-1}\varphi)^I X_I 
       \right)
  \right)
  \rangle
 \nonumber\\
  =
  \int_{\mathcal{M}}
  \hat{A}(\mathcal{M})
  ch\left(
     \iota_{x_0}^* (\mathcal{E}_{\lambda})|_{\mathcal{M}}
    \right)
  =
  \mathrm{ind}_{\iota_{x_0}^*( \mathcal{E}_{\lambda})|_{\mathcal{M}}}(D)
  .
\end{eqnarray}

\section{Discussion}
We expect that
 this method
 helps
 understanding of the emergence of geometries
 by condensations of Wilson loops
 \cite{nakatsu}.

\subsection*{Acknowledgment}

I thank Tudor Dimofte
 for his notification.
Y.~N. is supported 
 by JSPS-IHES-EPDI
 Fellowship.

\appendix
\section{Geometries of spaces of connections}
\label{sec;appendix}

In this appendix we describe
 geometries of spaces of connections.
In \ref{subsec;geom of a/g},
 we discuss geometries of a gauge-orbit space.
The instanton moduli space $\mathcal{M}$
 are considered as
 a subspace of the gauge-orbit space.
In \ref{subsec;vector bundle on a/g},
 we discuss vector bundles
 over the gauge-orbit space.
In this appendix,
 we consider elements of a Lie algebra 
 $\mathfrak{g}$
 as anti-Hermitian matrices to simplify notations.

\subsection{Geometries on $\mathcal{A}^*/\mathcal{G}_0$}
\label{subsec;geom of a/g}

We describe geometries of a gauge-orbit space,
 which are used
 the main part of this paper.
Almost all of the things in this subsection
 are written
 in \cite{geom of connection} 
 and \cite{donaldson kronheimer}.

For a real compact Lie group $G$,
 there is a $G$-principal bundle
 $P\rightarrow M_4$.
We denote the adjoint bundle by
 $AdP:= P\times_{ad} \mathfrak{g}$.
In general,
 for a vector bundle 
 $\mathcal{V}\rightarrow M_4$,
 let
 $\Omega^r(M_4,\mathcal{V})$
 be the space of $\mathcal{V}$-valued $r$-forms:
\begin{eqnarray}
 \Omega^r(M_4,\mathcal{V})
  :=
  \Gamma(M_4,\wedge^r T^*M_4\otimes \mathcal{V})
  .
\end{eqnarray}
The space of connection
 of $AdP$ are denoted by $\mathcal{A}$.
The space $\mathcal{A}$ is
 an affine space
 modeled on $\Omega^1(M_4,AdP)$.

Any vector fields on $\mathcal{A}$
 are described as follows:
\begin{eqnarray}
 \int_{M_4} dx\,
  \alpha_\mu^I(x;A)
 \frac{\delta}{\delta A_\mu^I(x)}
 \in
 T_A\mathcal{A},
\end{eqnarray}
 where
 $\alpha:\mathcal{A}\rightarrow\Omega^1(M_4,AdP)$.
From now on,
 to simplify notations
 we will omit the integral symbol
 and the functional derivative
 in vector fields.

The metric $g_{\mathcal{A}}$ on $\mathcal{A}$
 are defined as follows:
\begin{eqnarray}
 g_{\mathcal{A}}
  (\alpha,\beta)
  :=
  \int_{M_4}
  \mathrm{Tr}
  \left(
   \alpha \wedge
   * \beta
  \right)
  ,
  \hspace{5mm}
  \mbox{for $\alpha,\beta \in \Gamma(\mathcal{A},T\mathcal{A})$.}
\end{eqnarray}

Let $\mathcal{G}$
 and $Lie(\mathcal{G})=\Gamma(M_4,AdP)$
 be the gauge transformation group
 and its Lie algebra.
An element
 $\gamma\in\mathcal{G}$ acts on $A\in\mathcal{A}$
 as follows:
\begin{eqnarray}
 \gamma \cdot A
  =
  \gamma A \gamma^{-1}
  +\gamma d \gamma^{-1}
  .
\end{eqnarray}
Similarly, $\Lambda\in Lie (\mathcal{G})$ acts on
  $A\in\mathcal{A}$ as follows:
\begin{eqnarray}
 \Lambda \cdot A
  =
  A - d_A\Lambda
  ,
\end{eqnarray}
where $d_A:= d+ad(A)$.
Then, the vector field
 $\Lambda_{\mathcal{A}}\in\Gamma(\mathcal{A},T\mathcal{A})$
 corresponding to this action
 becomes as follows:
\begin{eqnarray}
 \Lambda_{\mathcal{A}}|_{A}
  =
  d_A \Lambda.
\end{eqnarray}

We denote the stabilizer group of $A\in\mathcal{A}$
 by $Stab(A)$:
 $Stab(A):=  \{  \gamma\in\mathcal{G}|\gamma\cdot A=A\}$.
A connection $A$ is called irreducible
 when the stabilizer group is the center of $G$:
\begin{eqnarray}
  Stab(A)=C(G)
  .
\end{eqnarray}
The space of irreducible connections
 is denoted by $\mathcal{A}^*$.
We consider the following subgroup 
 of $\mathcal{G}$:
\begin{eqnarray}
 \mathcal{G}_0
  =
  \{
  \gamma\in\mathcal{G}
  |
  \gamma(\infty)= e
  \}.
\end{eqnarray}
Since
 the stabilizer group
 of $A\in\mathcal{A}^*$
 is a subgroup of $G$
 and
 $\mathcal{G}_0$
 does not contains the global gauge transformation
 $G$,
 $\mathcal{G}_0$
 action on $\mathcal{A}$ is free. 
Therefore,
 the space $\mathcal{A}^*$
 can be considered
 as a $\mathcal{G}_0$-principal bundle:
 $\mathcal{A}^*\rightarrow \mathcal{A}^*/\mathcal{G}_0$.

We can consider a connection
 of $\mathcal{A}^*\rightarrow \mathcal{A}^*/\mathcal{G}_0$.
The vertical vectors
 are defined as vectors
 caused by the gauge transformations:
\begin{eqnarray}
 V_A\mathcal{A}^*
  :=
  \mathrm{Im}(d_A)
  .
\end{eqnarray}
The horizontal vectors
 are defined as
 the orthogonal vectors
 of the vertical vectors
 with respect to the metric 
 $g_{\mathcal{A}}$:
\begin{eqnarray}
 H_A\mathcal{A}
  :=
  \mathrm{Ker}(d_A^*)
  .
\end{eqnarray}
We denote the projection operator
 from tangent vectors
 to horizontal vectors
 by $P_H$:
\begin{eqnarray}
 P_H|_A
  =
  1-d_A \frac{1}{d_A^*d_A}d_A^*
  .
\end{eqnarray}
According to the decomposition of
 tangent vectors,
 the connection
 one-form
$\Theta\in \Gamma(\mathcal{A}^*,T^*\mathcal{A}\otimes
\Omega^1(M_4,AdP))$
 is given by the following equation:
\begin{eqnarray}
 \Theta|_A
  =
  \int_{M_4} dy\,
  G_A(\cdot,y)
  d_A^*|_y
  d_{\mathcal{A}}A(y)
  ,
\end{eqnarray}
where $G_A(x,y)$
 is the Green function of
 the Laplacian $d_A^* d_A$
 on $\Omega^0(M_4,AdP)$.

We identify the tangent vector space
 at $[A]\in\mathcal{A}^*/\mathcal{G}_0$
 and the horizontal vector space
 at $A\in\mathcal{A}^*$:
 $T_{[A]}\mathcal{A}^*/\mathcal{G}_0\simeq T_A\mathcal{A}^*$.
A metric
 $g_{\mathcal{A}^*/\mathcal{G}_0}$
 of $\mathcal{A}^*/\mathcal{G}_0$
 is defined by the following equation:
\begin{eqnarray}
g_{\mathcal{A}^*/\mathcal{G}_0}
 ([\alpha],[\beta])([A])
 :=
 g_{\mathcal{A}^*}
 (P_H\alpha,P_H\beta)(A)
 ,
 \hspace{5mm}
 \mbox{for $\alpha,\beta\in T_A\mathcal{A}^*$.}
\end{eqnarray}

We take an open set $U\subset\mathcal{A}^*/\mathcal{G}_0$,
 fix a local section
 $\sigma\in\Gamma(U,\mathcal{A}^*)$
 and introduce a coordinate $\{m^i\}$ on $U$.
We write a vector $b_i$
 as
 an abbreviation
 of the following vector:
\begin{eqnarray}
 b_i(x,m)
  :=
  P_H \circ \sigma_*(\frac{\partial}{\partial m^i}|_m)
  .
  \label{eq;def of bi}
\end{eqnarray}
With these notations
 the metric
 $g_{\mathcal{A}^*/\mathcal{G}_0}$
 is written as follows:
\begin{eqnarray}
 G_{ij}(m)
  =
  \int_{x\in M_4}
  \mathrm{Tr}_{\mathfrak{g}}
  \left(
   b_i\wedge* b_j
  \right)(x,m)
  .
  \label{eq;Gij}
\end{eqnarray}
Furthermore,
 we write the local connection
 one-form on $U$ 
 by
 $\epsilon\in\Gamma(U,T^*U\otimes Lie(\mathcal{G}))$:
\begin{eqnarray}
 \epsilon
 =
 \epsilon_i dm^i
  &:=&
  \sigma^*(\Theta).
\end{eqnarray}

From the decomposition
 of the vector field
 $\sigma_*(\frac{\partial}{\partial m^i})$,
 we obtain the following equation:
\begin{eqnarray}
 \frac{\partial A}{\partial m^i}(x;m)
  =
  b_i(x,m)
  +d_A \epsilon_i(x,m)
  .
  \label{eq;del_i A}
\end{eqnarray}

If we regard
 the curvature $F_A=dA+A\wedge A$ 
 as a map
 from $\mathcal{A}$
 to $\Omega^2(M_4,AdP)$,
 the instanton moduli space
 $\mathcal{M}$
 are considered
 as a subspace
 of $\mathcal{A}^*/\mathcal{G}_0$
 determined by the condition
 $F^{(+)}=0$.
In this subspace we need to impose
 the following condition
 on the tangent vectors $b_i$:
\begin{eqnarray}
 (d_A b_i)^{(+)}=0.
\end{eqnarray}

\subsection{Vector bundles over $\mathcal{A}^*/\mathcal{G}_0$}
\label{subsec;vector bundle on a/g}

We consider a vector bundle 
 called the universal bundle
 over 
 $M_4\times \mathcal{A}^*/\mathcal{G}_0$
 \cite{universal bundle}.
Furthermore,
 we give
 a definition
 of a certain vector bundle
 over $\mathcal{A}^*/\mathcal{G}_0$
 which is needed to the discussion of this paper.

We consider a projection 
 $\pi_1:M_4\times \mathcal{A}^*\rightarrow M_4$
 and a pullback bundle
 $\pi_1^*(AdP)\rightarrow M_4\times \mathcal{A}^*$.
At a point $(x,A) \in M_4\times \mathcal{A}^*$,
 we define a covariant derivative of the pullback bundle
 by
 $d_A$ along $M_4$ direction
 and
 $d_{\mathcal{A}}$ along  $\mathcal{A}^*$
 direction.
The group $\mathcal{G}_0$
 cause a bundle map on
 $\pi_1^*(AdP)$.
Then we can define a vector bundle 
$\mathcal{E}$
 called the universal bundle
 as follows:
\begin{eqnarray}
 \pi_1^*(AdP) &\substack{/\mathcal{G}_0\\ \longrightarrow}
&\mathcal{E}\\
\downarrow & &\downarrow \\
 M_4\times \mathcal{A}^*
  &\substack{/\mathcal{G}_0\\ \longrightarrow}&
M_4\times \mathcal{A}^*/\mathcal{G}_0
.
\end{eqnarray}
The covariant derivative
 on 
 $\mathcal{E}$
 is given by the following equation:
\begin{eqnarray}
 d^{\mathcal{E}}_{A+\epsilon}
  =
  d+A +
  dm^i\wedge(\partial_i +\epsilon_i)
  .
\end{eqnarray}
The curvature becomes as follows
 \cite{donaldson kronheimer}:
\begin{eqnarray}
 F^{\mathcal{E}}(\partial_\mu,\partial_\nu)
  (x,m)
  &=&
  \partial_\mu A_\nu
  -\partial_\nu A_\mu
  +\left[A_\mu,A_\nu\right]
  ,
  \\
 F^{\mathcal{E}}(\partial_i,\partial_\mu)
  (x,m)
  &=&
  b_{i\mu}
  ,
  \\
 F^{\mathcal{E}}(\partial_i,\partial_j)
  (x,m)
  &=&
  -2
  \left(
  \frac{1}{d_A^*d_A}
  \left[b_i\asi^\mu, b_{j\mu}\right]
  \right)
  (x,m)
  .
  \label{eq;curv of univ}
\end{eqnarray}

We apply the construction
 of the universal bundle to 
 other associated vector bundles
 over $M_4$.
For a given highest representation
 $(\pi_\lambda,V_\lambda)$ of $G$,
 we define a vector bundle 
 $E_\lambda:= P\times_{\pi_\lambda} V_\lambda$.
By using the projection
 $\pi_1:M_4\times \mathcal{A}^*\rightarrow M_4$,
 we define a pullback bundle
 $\pi_1^*(E_{\lambda})\rightarrow M_4\times \mathcal{A}^*$.
By dividing the vector bundle $\pi_1^*(E_{\lambda})$
 by $\mathcal{G}_0$,
 we obtain a vector bundle
 $\mathcal{E}_{\lambda}\rightarrow M_4\times
 \mathcal{A}^*/\mathcal{G}_0$,
 whose typical fiber is $V_\lambda$.
Furthermore,
 by using the following inclusion map $\iota_{x_0}$
\begin{eqnarray}
 \iota_{x_0}:
  \mathcal{A}^*/\mathcal{G}_0
  &\rightarrow&
  M_4\times \mathcal{A}^*/\mathcal{G}_0,
  \\
 (m)&\mapsto&(x_0,m),
\end{eqnarray}
we define a vector bundle
 $\iota_{x_0}^* \mathcal{E}_{\lambda}$
 over $\mathcal{A}^*/\mathcal{G}_0$.
The curvature
 of
 $\iota_{x_0}^*( \mathcal{E}_{\lambda})$
 at $m\in\mathcal{A}^*/\mathcal{G}_0$
 is given by the following equation:
\begin{eqnarray}
 \frac{1}{2}
 \pi_{\lambda}
  \left(
   F^{\mathcal{E}}(\partial_i,\partial_j)
   (x_0,m)
  \right)
  dm^i\wedge dm^j .
  \label{eq;curv of iota e}
\end{eqnarray}


\end{document}